\documentclass[12pt]{article}
\usepackage[margin=2.5cm]{geometry}  % add also "show frame" options  
\usepackage{graphicx,psfrag,epsf}
\usepackage{enumerate}
\usepackage{url} % not crucial - just used below for the URL 
\usepackage{float}
\usepackage{algorithm}
\usepackage{algorithmic}
\usepackage{amsmath}
\usepackage{amssymb}
\usepackage{graphicx}
\usepackage{subcaption}
\usepackage{natbib}
\usepackage{lineno}
\usepackage{dcolumn}
\usepackage{multirow}
\usepackage{units}
\usepackage{booktabs}
\usepackage{colortbl}
\usepackage{rotating}
\usepackage{epstopdf}
\usepackage[table]{xcolor}  
\newcolumntype{.}{D{.}{.}{-1}}
\newcolumntype{d}[1]{D{.}{.}{#1}}
\def\Z{\mbox{Z\hspace{-0.30em}Z}}

\usepackage{hyperref}
\hypersetup{
	colorlinks=true,
	citecolor=blue,
}

\usepackage{rotating}
\usepackage{epsfig}%
\usepackage{epstopdf}

\def\bTheta{\mbox{\boldmath $\Theta$}}

\def\L{\mbox{L}}
\def\bPsi{\mbox{\boldmath $\Psi$}}

\def\bgamma{\mbox{\boldmath $\gamma$}}

\def\balpha{\mbox{\boldmath $\alpha$}}

\def\bf{\mbox{\boldmath $f$}}

\def\I{\mbox{I}}

\def\1{\mbox{1}}

\def\E{\mbox{E}}

\def\T{\mbox{T}}

\def\balpha{\mbox{\boldmath $\alpha$}}

\def\bgamma{\mbox{\boldmath $\gamma$}}

\def\bdelta{\mbox{\boldmath $\delta$}}

\def\bveps{\mbox{\boldmath $\varepsilon$}}

\def\bTheta{\mbox{\boldmath $\Theta$}}
\def\bTheta{\mathbf{\Theta}}

\def\bPhi{\mathbf{\Phi}}
\def\bPsi{\mathbf{\Psi}}
\def\balpha{\mbox{\boldmath $\alpha$}}

\def\bgamma{\mbox{\boldmath $\gamma$}}

\def\bdelta{\mbox{\boldmath $\delta$}}

\def\bveps{\mbox{\boldmath $\varepsilon$}}

\def\bTheta{\mbox{\boldmath $\Theta$}}

\def\bTheta{\mathbf{\Theta}}

\def\bPhi{\mathbf{\Phi}}
\def\bPsi{\mathbf{\Psi}}

\def\bu{\mathbf{u}}
\def\by{\mathbf{y}}

\def\0{\mbox{\bf{0}}}

\def\caliW{\mbox{\boldmath ${\cal W}$}}

\def\calI{\mbox{\boldmath ${\cal I}$}}
\def\E{\mbox{E}}

\def\bu{\mathbf{u}}

\def\by{\mathbf{y}}

\def\0{\mbox{\bf{0}}}

\def\bu{\mathbf{u}}
\def\by{\mathbf{y}}

\def\0{\mbox{\bf{0}}}

\def\E{\mbox{E}}

\def\T{\mbox{T}}

\def\L{\mbox{L}}

\def\Z{\mbox{Z}}

\def\I{\mbox{I}}

\def\bkR{{\rm I\kern-.17em R}}

\def \1n{1\hskip -3pt \mbox{N}}

\newfont{\bbf}{cmbx12 scaled 1435}
\begin{document}

\setlength{\baselineskip}{.26in}
\thispagestyle{empty}
\renewcommand{\thefootnote}{\fnsymbol{footnote}}
\vspace*{0cm}
\begin{center}

\setlength{\baselineskip}{.32in}
{\bbf Sequential Monte Carlo for Noncausal Processes}\\

\vspace{0.5in}

\large{Gianluca Cubadda}\footnote{Department of Economics and Finance, University of Rome 'Tor Vergata', Italy:\\gianluca.cubadda@uniroma2.it} \large{Francesco Giancaterini}\footnote{Centre for Economic and International Studies, University of Rome 'Tor Vergata', Italy:\\francesco.giancaterini@uniroma2.it} \large{Stefano Grassi}\footnote{Department of Economics and Finance, University of Rome 'Tor Vergata', Italy:\\stefano.grassi@uniroma2.it}\\

\setlength{\baselineskip}{.26in}

\vspace{0.4in}

\today\\

\medskip

\vspace{0.3in}
\begin{minipage}[t]{12cm}
\small
\begin{center}
Abstract \\
\end{center}

%Mixed causal and noncausal models have gained popularity in recent years in finance and macroeconomics.
%however Bayesian inference for those models is still challenging. 

This paper proposes a Sequential Monte Carlo approach for the Bayesian estimation of mixed causal and noncausal models. Unlike previous Bayesian estimation methods developed for these models, Sequential Monte Carlo offers extensive parallelization opportunities, significantly reducing estimation time and mitigating the risk of becoming trapped in local minima, a common issue in noncausal processes. Simulation studies demonstrate the strong ability of the algorithm to produce accurate estimates and correctly identify the process. In particular, we propose a novel identification methodology that leverages the Marginal Data Density and the Bayesian Information Criterion. Unlike previous studies, this methodology determines not only the causal and noncausal polynomial orders but also the error term distribution that best fits the data. Finally, Sequential Monte Carlo is applied to a bivariate process containing S$\&$P Europe 350 ESG Index and Brent crude oil prices.

\bigskip

\end{minipage}

\end{center}
\renewcommand{\thefootnote}{\arabic{footnote}}

\section{Introduction}
Since their inception by \cite{breid1991maximum}, Mixed causal-noncausal Auto-Regressive models (MAR) have gained significant popularity due to their ability to capture nonlinear dynamics in time series. Specifically, unlike conventional tests that detect explosive behaviors through unit roots (\citealp{phillips2011explosive}, \citealp{phillips2015testing}), MARs treat the variable under investigation as stationary and interpret bubble patterns as an intrinsic part of its dynamics, see, among others, \cite{fries2019mixed}, \cite{gourieroux2013explosive}, \cite{cavaliere2020bootstrapping} and \cite{hecq2021forecasting}. Moreover, MAR models are also an important tool in macroeconomics due to their ability to model expectations and their impact on the variable under investigation, see \cite{lanne2011noncausal}. Finally, they are also useful for testing nonfundamentalness, see \cite{alessi2011non} and \cite{lanne2013noncausal}.

Over time, both parametric and semiparametric estimators have been developed for MARs. An example of a semiparametric estimator is the Generalized Covariance (Gcov), which aims to minimize a portmanteau-type objective function involving the autocovariances of linear and nonlinear transformations of model errors (see \citealp{gourieroux2017noncausal} and \citealp{gourieroux2023generalized}).
Another semiparametric estimator is the one proposed by \cite{hecq2022spectral}, which exploits information from higher-order cumulants by combining the spectrum and the bispectrum in a minimum distance estimation. 
In the parametric framework, the maximum likelihood estimator for univariate models was introduced by \cite{breid1991maximum} and subsequently updated by \cite{lanne2011noncausal}. For multivariate models, the maximum likelihood estimator was introduced by \cite{lanne2013noncausal} and \cite{davis2020noncausal}. 

An alternative approach to estimating MARs is the Bayesian method introduced by \cite{lanne2012bayesian} for univariate models, which is based on the Markov Chain Monte Carlo (MCMC) technique. The multivariate extension, discussed in \cite{lanne2016noncausal}, takes advantage of the Metropolis-within-Gibbs sampler.

This paper aims to extend and adapt the Bayesian estimation method based on the sequential Monte Carlo algorithm (SMC), as illustrated in \cite{herbst2014sequential} and \cite{bognanni2018sequential}, to MARs. SMC offers an attractive alternative to MCMC for different reasons. First, this method is highly adaptable and can be easily implemented under different assumptions about the error term, as long as the posterior kernel can be evaluated point-wise. Second, it allows for fast initialization with random draws from the prior, unlike MCMC, which often requires a slow mode search.
Third, SMC generates the Marginal Data Density (MDD) of the model as a byproduct, simplifying the calculation of Bayes factors without additional computational effort. 
Finally, SMC is parallelizable (see \citealp{durham2014adaptive}), which not only significantly reduces the estimation time but also helps with objective functions containing local minima, a common issue in MARs. Specifically, \cite{bec2020mixed} highlights local minima in the objective function of the maximum likelihood estimator; \cite{hecq2022spectral} in the spectral estimation function; \cite{cubadda2024optimization} in the GCov objective function; \cite{lanne2016noncausal} notes non-elliptical shapes, such as skewness and multimodality, in marginal posterior distributions.

The existing literature on noncausal models has explored Bayesian estimation only under the assumption that the error term follows a Student-$t$ distribution; see, among others, \cite{lanne2012bayesian}, \cite{lanne2016noncausal}, \cite{nyberg2014forecasting} and \cite{moussa2023identifying}. However, this assumption may be too restrictive in some cases, and alternative distributions may provide a better fit. For example, as shown in \cite{gourieroux2017local}, the Cauchy distribution can more accurately capture extreme events, such as market crashes, while the Skewed $t$ distribution is better suited to model asymmetric patterns in financial time series, as discussed in \cite{proietti2023peaks}. Since the distribution of the error term is usually unknown to the researcher, we propose to exploit the adaptability of SMC to distributions and introduce an identification methodology, based on the MDD and the Bayesian Information Criterion (BIC), that not only determines the causal and noncausal polynomial orders but also identifies the error term distribution that best fits. To the best of our knowledge, none of the already existing methodologies allows for simultaneously performing both model selection and identification of the error distribution.

The rest of the paper is organized as follows. Section \ref{sec:Model} reviews the univariate and multivariate MAR. Section \ref{sec:BayesianEst} introduces the SMC for MARs. In Section \ref{sec:MonteCarlo} the ability of SMC for estimation and model identification is investigated by Monte Carlo exercises. Section \ref{sec:Empirical} reports the empirical application. Finally, Section \ref{sec:Concl} draws some conclusions. Further results and derivations are reported in the Appendices.

\section{Theory}
\label{sec:Model}
\subsection{Univariate Mixed Causal-Noncausal Autoregressive Processes}
\label{subsec:Univariate}
Univariate MAR processes, also denoted as MAR($r,s$), have the following representation:
\begin{equation}
    \psi(\L)\phi(\L^{-1})y_t = u_t,
    \label{eq:MAR}
\end{equation}
where $\psi(\L) = \left(1 - \psi_1 \L - \dots - \psi_r \L^r\right)$ is the causal polynomial of order $r$; $\phi(\L^{-1}) = \left(1 - \phi_1 \L^{-1} - \dots - \phi_s \L^{-s}\right)$ is the noncausal polynomial of order \(s\); and $p$ is defined as $p = r + s$. Moreover, both polynomials have roots outside the unit circle:
\begin{equation*}
\psi (z) \neq 0 \text{ for } |z|\leq 1 \hspace{0.25cm}\text{    and    }\hspace{0.25cm}
\phi (z) \neq 0 \text{ for } |z|\leq 1.
\end{equation*}

Due to the noncausal polynomial, the process in \eqref{eq:MAR} is capable of capturing nonlinear dynamics, including local trends (bubbles) and conditional heteroscedasticity (see \citealp{hencic2015noncausal}, \citealp{hecq2016identification} and \citealp{gourieroux2018misspecification}). Finally, to successfully identify these models, it is well known that $u_t$ must be independent and identically distributed ($i.i.d.$) and non-Gaussian, see \cite{breid1991maximum} and \cite{lanne2011noncausal}.\\
\indent The series $y_t$ in \eqref{eq:MAR} admits a two-sided moving average representation:
\begin{equation*}
y_t = \sum_{j=-\infty}^{\infty} \theta_j u_{t+j},
\end{equation*}
where $\theta_0$ is equal to 1. If the process is purely causal ($s=0$) or purely noncausal ($r=0$), then $\theta_j = 0$ for $j > 0$ or for $j < 0$, see \cite{gourieroux2015uniqueness}.

\subsection{Multivariate mixed causal and noncausal processes}
\label{subsec:Multivariate}
Vector MAR processes, denoted as VMAR($r$,$s$) were introduced by \cite{lanne2013noncausal} and are given by:
\begin{equation}
    \Psi(\L)\Phi(\L^{-1})\by_t = \bu_t,
    \label{eq:VMAR}
\end{equation}
where: $\by_t = \left(y_{1,t}, \dots , y_{n,t}\right)^{\prime}$ and $\bu_t = \left(u_{1,t}, \dots , u_{n,t}\right)^{\prime}$ are two $n \times 1$ vectors; $\Phi(\L^{-1}) = \I_n - \Phi_1\L^{-1} - \dots - \Phi_s \L^{-s} $ is a $n \times n$ noncausal polynomial of order $s$; and $\Psi(\L) = \I_n - \Psi_1\L - \dots - \Psi_r\L^{r}$ is a $n \times n$ causal polynomial of order $r$. Finally, $p$ is defined as $p = r + s$. As in the univariate framework, both the causal and noncausal polynomials have their roots outside the unit circle:
\begin{equation*}
|\Psi (z)| \neq 0 \text{ for } |z|\leq 1 \hspace{0.25cm}\text{    and    }\hspace{0.25cm}
|\Phi (z)| \neq 0 \text{ for } |z|\leq 1,
\end{equation*}
and the identification of the noncausal component requires two conditions: the error term must be $i.i.d.$ and non-Gaussian, see \cite{gourieroux2017noncausal} and \cite{cubadda2023detecting}.\\
\indent Similar to Section \ref{subsec:Univariate}, the process in \eqref{eq:VMAR} exhibits a two-sided moving average specification: 
\begin{equation*}
\by_t = \sum_{j=-\infty}^{\infty} \Theta_j \bu_{t+j},
\end{equation*}
where $\Theta_0$ is equal to a  $n \times n$ identity matrix. If the process is purely causal ($s=0$), or purely noncausal ($r=0$), then $\Theta_j$ is a null matrix for $j>0$ or for $j<0$; see, among others, \cite{lanne2013noncausal}.

%\indent It is important to note that at the multivariate level, specification VMAR($r,s$) does not cover all possible forms of noncausality. Indeed, 
As shown in \cite{gourieroux2017noncausal} and \cite{davis2020noncausal}, a potential alternative mixed model specification is given by: 
\begin{equation} 
    \by_t = \sum_{j=1}^{p} \Pi_j \by_{t-j} + \bveps_t, 
    \label{eq:VAR} 
\end{equation} 
where the autoregressive polynomial $|\Pi(z)| = |\I_n - \sum_{j=1}^{p} \Pi_j z^j|$ can have roots both inside and outside the unit circle. Specifically, the process defined in \eqref{eq:VAR} is considered purely causal if all roots lie outside the unit circle, purely noncausal if all roots lie inside the unit circle, and mixed causal and non-causal if the roots lie both inside and outside the unit circle. As highlighted in \cite{lanne2013noncausal}, specifications \eqref{eq:VMAR} and \eqref{eq:VAR} do not always overlap, and there are cases where the specification \eqref{eq:VMAR} cannot be embedded in \eqref{eq:VAR}, and vice versa. The question of whether there is a feasible specification that incorporates all or the majority of noncausal VAR processes is intriguing, but it is beyond the scope of this paper. Here, we focus exclusively on the VMAR($r,s$) specification. This is because, in addition to encompassing the univariate specification in \eqref{eq:MAR} as a special case, it aligns with the representation used in \cite{lanne2016noncausal}. Finally, the VMAR($r,s$) representation allows for a neater interpretation of the results by explicitly separating the causal and noncausal components.

\section{Bayesian Estimation}
\label{sec:BayesianEst}

\subsection{The Bayesian Estimation of VMAR Processes}
\label{subsec:BayesianVMAR}

The Bayesian analysis of mixed models has been examined by \cite{lanne2012bayesian} for univariate models and by \cite{lanne2016noncausal} for multivariate models. We focus on the multivariate framework, as the process in \eqref{eq:MAR} is a special case of the process in \eqref{eq:VMAR} when $n=1$; for the univariate example, see Appendix \ref{appendix:univariate}. 

Let $\by_t$ be generated by \eqref{eq:VMAR}, and define $\by=\left[  \by_{1},\dots,\by_{T}\right]  ^{\prime}$, where $T$ denotes the sample size. We also define $\bTheta_1 = \left[\text{vec}(\Psi_1)^{\prime}, \dots, \text{vec}(\Psi_s)^{\prime}, \text{vec}(\Phi_1)^{\prime}, \dots, \text{vec}(\Phi_r)^{\prime}\right] ^{\prime}$) and $\bTheta_2$ a column vector that contains all the parameters that depend on the distribution of $\bu_t$ (e.g., degrees of freedom, the vector half operator of the scale matrix, etc.). By defining the $k$-vector $\bTheta = [\bTheta_1^{\prime}, \bTheta_2^{\prime}]^{\prime}$, Bayes' theorem computes posterior probabilities by combining prior beliefs with observed data: 
\begin{equation*} 
    p(\bTheta|\by) = \frac{p(\by|\bTheta) p(\bTheta)}{p(\by)} = \frac{p(\by|\bTheta) p(\bTheta)}{\int p(\by|\bTheta) p(\bTheta) d\bTheta}, 
\end{equation*}
where $p(\bTheta|\by)$ denotes the posterior probability, $p(\by|\bTheta)$ represents the likelihood function, $p(\bTheta)$ indicates the prior probability distribution of the parameters, and $p(\by)$ represents the MDD, or marginal likelihood, which is important for model selection and prediction.

To compute $p(\bTheta|\by)$, \cite{lanne2012bayesian} and \citet{lanne2016noncausal} employed a tailored Gibbs sampling algorithm. For MDD, p($\by$), \citet{lanne2012bayesian} used the method introduced by \citet{geweke2005bayesian}, while \citet{lanne2016noncausal} adopted a technique similar to MitISEM. In this paper, we propose to compute the posterior distribution and MDD for both the univariate and multivariate frameworks using a different approach, namely the SMC algorithm, as described in Section \ref{subsec:SMC}. As mentioned above, SMC enables the calculation of MDD as a simple by-product, without the need for additional processing. It also mitigates the risk of local minima, a common problem in the posterior distributions of MARs (see \citealp{lanne2016noncausal} and \citealp{bognanni2018sequential}). Finally, SMC is highly adaptable to a wide range of error term distributions, provided that the posterior kernel can be evaluated point-wise.

\subsection{Sequential Monte Carlo for VMAR processes}
\label{subsec:SMC}
The SMC algorithm proposed by \cite{herbst2014sequential} is based on Importance Sampling (IS). The IS approximates the target density $f(\cdot)$ using a different and easily to sample density $g(\cdot)$ called the candidate density:
\begin{equation*}
   \E_{\pi} [ h(\bTheta)] = \int h(\bTheta)\pi(\bTheta) d\bTheta=\frac{1}{\Z}\int h(\bTheta)w(\bTheta)g(\bTheta)d\bTheta, 
\end{equation*}
where $h:\bTheta\rightarrow\mathbb{R}^k$, $\pi(\bTheta)=p(\bTheta|\by)$, $f(\bTheta)=p(\by|\bTheta)p(\bTheta)$, $\Z=p(\by)$, and $w(\bTheta)=f(\bTheta)/{g(\bTheta)}$ are the importance sample weights. If $\bTheta^{i^{\text{i.i.d.}}} \sim g(\theta)$, for $ i = 1, \dots, P $, then, under suitable regularity conditions (\citealp{geweke1989bayesian}), the Monte Carlo estimate is given by:
\begin{equation*}
    \bar{h} = \frac{1}{P} \sum_{i=1}^{P} h(\bTheta_i) \tilde{W}_i, \quad \text{where} \quad \tilde{W}_i = \frac{w(\theta_i)}{\frac{1}{P} \sum_{j=1}^{P} w(\theta_j)}.
\end{equation*}
As $ P $ goes to infinity, it converges almost surely (a.s.) to $\E_\pi[h(\bTheta)]$, see \cite{bognanni2018sequential}. Each $\tilde{W}_i$, referred to as a (normalized) importance weight (IW), is assigned to an associated parameter $\bTheta_i$. We refer to the pair $ (\bTheta_i, \tilde{W}_i) $ as a particle. The collection of particles $ \{ (\bTheta_i, \tilde{W}_i) \}_{i=1}^{P} $ forms a discrete distribution that provides an approximation of $ \pi(\bTheta)$. The accuracy of the approximation is determined by the divergence between $ g(\cdot) $ and $ f(\cdot) $. However, constructing a suitable proposal distribution $g(\cdot)$ is challenging, especially when little is known about the shape of $f(\cdot)$. SMC offers a clever solution to this problem by iteratively constructing particle approximations to a sequence of distributions. Starting from the prior distribution, a sequence of bridge distributions that gradually incorporates more likelihood information is constructed. This procedure is repeated until the full likelihood has been assimilated in the posterior distribution, more precisely: 
\begin{equation*}
    \pi_m(\bTheta)=\frac{f_m(\bTheta)}{\Z_m}=\frac{[p(\by|\bTheta)]^{\rho_m}p(\bTheta)}{\int [p(\by|\bTheta)]^{\rho_m}p(\bTheta) d\bTheta}, \ \ \ m=1, \dots, M,
\end{equation*}
where $M$ are the number of stages that depend on the complexity of the problem at hand, and the values of the tempering parameter $\rho_m$ are given by an increasing sequence of values such that $\rho_1=0$ and $\rho_{M}=1$. Notice that $\rho_m$ is constructed so that more and more likelihood information is incorporated in the posterior as $m$ increases. As in \cite{herbst2014sequential} and \cite{bognanni2018sequential}, we define it as: 
\begin{equation}
\label{eq:rho}
    \rho_m = \left(\frac{m-1}{M -1}\right)^\lambda,
\end{equation}
where the hyper-parameter $\lambda$, which is greater than 0, determines the rate at which the likelihood information is incorporated into the sampler. In particular, when $\lambda = 1$ the schedule is linear, and each stage contributes equally. When $\lambda > 1$, smaller increments in likelihood are prioritized in the initial stages, with larger increments occurring in the later stages. Conversely, when $\lambda < 1$ larger increments in the early stages are prioritized, with smaller increments in later stages.

A key drawback of IS is weight degeneracy, which occurs when the target distribution is sharp relative to the proposal, resulting in most IWs approaching zero. This leads to an insufficient number of representative samples and is closely related to the curse of dimensionality. Specifically, the issue arises from the extreme values the likelihood takes on in high-dimensional spaces. SMC mitigates this challenge through a tempering approach that progressively incorporates the likelihood.

Finally, following \cite{bognanni2018sequential} the MDD take the following form:
\begin{equation}
    \text{MDD}= \prod_{m=1}^{M} \left( \sum_{i=1}^{P} \tilde{w}_n^i \right),
    \label{eq:MDD}
 \end{equation}
 where $P$ are the IS draws, and $\tilde{w}_n^i$ are the 
the incremental and normalized weights, obtained by:
$$
\tilde{w}_m^i = [ p(\by|\bTheta_{m-1}^i)]^{\rho_m-\rho_{m-1}}.
$$
See Appendix \ref{appenidix_B} for a detailed description of the SMC algorithm and MDD calculation.

\section{Monte Carlo Analysis}
\label{sec:MonteCarlo}
The performance of the SMC algorithm for VMAR($r,s$) models is evaluated using three Monte Carlo exercises. In each exercise, we set $r=s=1$, $n=2$ and the error term is generated from one of the following multivariate distributions: Cauchy, Student-$t$, and Skewed-$t$. For each Data Generating Process (DGP), we estimate several VMAR($r,s$) models using these three error distributions and various combinations of polynomial orders $r$ and $s$ to examine how often the algorithm correctly identifies them. Furthermore, the accuracy of the algorithm in estimating true population parameters is evaluated under the assumption that both the error distribution and the polynomial orders are known.

Sections \ref{subsec:priors} and \ref{subsec:lik} describe the priors and likelihood functions used in the Monte Carlo exercises, while Section \ref{subsec:MCresults} outlines the Monte Carlo settings and presents the results.

\subsection{Priors}\label{subsec:priors}
As discussed in Section \ref{subsec:BayesianVMAR}, $\bTheta=[\bTheta_1^{\prime}, \bTheta_2^{\prime}] ^{\prime}$ indicates the prior probability of the VMAR($r,s$) parameters. Specifically, $\bTheta_1$, which collects the elements of the VMAR($r,s$) coefficient matrices, remains unchanged regardless of the error term distribution, whereas $\bTheta_2$ varies with the density distribution of $\bu_t$ since it includes the vectorization of the scale matrix $\Sigma$, which is shared by the Student-$t$, Cauchy, and Skewed-$t$ distributions; the degrees of freedom ($\nu$) for the Student-$t$ and Skewed-$t$ distributions; and the vector containing the skewness parameters ($\balpha=\left[\alpha_1, \dots , \alpha_n\right]^{\prime}$) for the Skewed-$t$ distribution.

For $\bPsi$ and $\bPhi$ we use Multivariate Normal ($ \mathcal{MN}(\cdot, \cdot)$) priors:
\begin{equation}
    \bPsi \sim \mathcal{MN}\left(0, \bgamma \text{I}_{n^2r}\right)\calI(\bPsi),
        \hspace{1.5cm}  \bPhi \sim \mathcal{MN}\left(0, \bdelta \text{I}_{n^2s}\right)\calI(\bPhi),
    \label{eq:causal}
\end{equation}
%and:
%\begin{equation}
%    \bPhi \sim \mathcal{N}\left(0, \bgamma \text{I}_{n^2s}\right)\calI(\bPhi),
 %   \label{eq:noncausal}
%\end{equation}
where $\calI(\bPsi)$ and $\calI(\bPhi)$ are indicator functions equal to unity in the stationary region. As in \cite{lanne2016noncausal}, the parameters $\bgamma$ and $\bdelta$ decrease as the number of leads and lags increases, implying the use of Minnesota-type priors. We set $\bgamma = 2/i$ for $i = 1, \dots, r$ and $\bdelta = 2/q$ for $q = 1, \dots, s$. Thus, we consider more diffuse priors compared to \cite{lanne2016noncausal}, who set $\bgamma = 1/i$ and $\bdelta = 1/q$.

The prior distribution for the scale matrix, which is common to all the error distributions, is an Inverse Wishart ($\caliW^{-1}(\cdot, \cdot)$):
\begin{equation}
    \Sigma \sim \caliW^{-1}\left(\Psi_0, \tilde{\nu}_{\Sigma}\right),
    \label{eq:scale}
\end{equation}
where a diffuse prior is adopted by setting $\Psi_0 =5\I_n$ for the scale matrix and $\tilde{\nu}_{\Sigma} = 3$ for the degrees of freedom

Finally, we use Exponential prior ($\mathcal{E}(\cdot)$) for the degrees of freedom ($\nu$) and $\mathcal{MN}(\cdot, \cdot)$ for the skewness parameter $\alpha$:
\begin{equation}
    \nu \sim \mathcal{E}(\nu_0),  \hspace{2.5cm}  \balpha \sim \mathcal{MN}(0,\kappa\I_n),
    \label{eq:df}
\end{equation}
%and:
%\begin{equation}
%    \alpha \sim \mathcal{N}(0,q\text{I}_n),
%    \label{eq:alpha}
%\end{equation}
%respectively. 
where, we set $\nu_0=5$ and $\kappa=3$.

\subsection{Likelihood functions}\label{subsec:lik}
The second component of the Bayesian analysis described in Section \ref{subsec:BayesianVMAR} is the likelihood function $p(\by \mid \bTheta)$. We use the same (approximate) likelihood functions used in \cite{lanne2013noncausal} and \cite{lanne2016noncausal}, given by:
\begin{equation*}
    p(\by \mid \bTheta)=\sum_{t=r+1}^{T-s}p(\bu_t \mid \bTheta),
\end{equation*}
where $T$ is the sample size, and $p(\bu_t \mid \bTheta)$ varies according to the error term distribution considered. Specifically, for the Multivariate Student-$t$ we have:
\begin{equation}
    p^{Student-t} (\bu_t \mid \bTheta) = \sum_{t=r+1}^{T-s} \left[\frac{\Gamma\left(\frac{\nu + n}{2}\right)}{\Gamma\left(\frac{\nu}{2}\right) (\pi \nu)^{n/2} |\Sigma|^{1/2} }
    \left(1 + \frac{1}{\nu} \bu_t^\prime \Sigma^{-1} \bu_t\right)^{-\frac{\nu + n}{2}}\right],
    \label{eq:MST}
\end{equation}
where $\nu$ indicates the degrees of freedom of the error term, see \cite{lanne2013noncausal}. 

The Multivariate Skewed-$t$ distribution has the following likelihood function:
\begin{equation}
    p^{Skewed-t} (\bu_t \mid \bTheta) = \sum_{t=r+1}^{T-s} \left\{ 2 t_n(\bu_t , \nu)\T_1 \left[\alpha^{\prime}u_t \left(\frac{\nu+n}{\left(\bu_t^\prime \Sigma^{-1} \bu_t\right)+\nu}\right)^{1/2} , \nu+n \right] \right\},
    \label{eq:MSK}
\end{equation}
where $\alpha$ is an $n$-vector of shape parameters of the error term, $t_n(\bu_t,\nu)$ is the probability density function of an $n$-dimensional Student-$t$ random variable with $\nu$ degrees of freedom, and $\T_1[\cdot, \nu + n]$ is the cumulative distribution function of a scalar Student-$t$ distribution with $\nu + n$ degrees of freedom. 

The Cauchy distribution has the following likelihood function:
\begin{equation}
    p^{Cauchy} (\bu_t \mid \bTheta) = \sum_{t=r+1}^{T-s} \left[\frac{\Gamma\left(\frac{1+n}{2}\right)}{\Gamma\left(\frac{1}{2}\right)(\pi)^{n/2} |\Sigma|^{1/2} }
    \left(1 + \bu_t^\prime \Sigma^{-1} \bu_t\right)^{-\frac{1+n}{2}}\right],
    \label{eq:Cauchy}
\end{equation}
that is a special case of \eqref{eq:MST} when $\nu$ is fixed to 1.

\cite{azzalini2003distributions} underlines that when $\alpha$ is a zero vector, then equations \eqref{eq:MST} and \eqref{eq:MSK} coincide.  Moreover, \cite{lanne2016noncausal} considers the multivariate Student-$t$ only for cases where $\nu > 2$, which ensures a finite variance of the error term. We impose the same restriction on \eqref{eq:MST} and \eqref{eq:MSK}. Finally, in equation \eqref{eq:MST}, \eqref{eq:MSK} and \eqref{eq:Cauchy} it is common to lose the first $r$ and last $s$ observations as discussed in \cite{lanne2011noncausal, lanne2013noncausal}.

% ated to the univariate framework, see Appendix B.

\subsection{Monte Carlo experiments}
\label{subsec:MCresults}
As previously stated, we set $r = s = 1$, $n = 2$, and we assume that the error term follows three multivariate distributions: Cauchy, Student-$t$, and Skewed-$t$. We consider a sample size of $T = 150$ observations and $B = 200$ Monte Carlo replications. Relatively small values for $T$ and $B$ are chosen because increasing them would lead to excessive computational time for the Monte Carlo exercise.

For each DGP, we estimate twenty-one models in each replication: VMAR($1,0$), VMAR($0,1$), VMAR($2,0$), VMAR($1,1$), VMAR($0,2$), VMAR($2,1$), and VMAR($1,2$) across the Cauchy, Student-$t$, and Skewed-$t$ distributions. Finally, in the SMC algorithm, we set the number of particles $P = 10000$, the rate at which the likelihood information is incorporated into the sample $\lambda = 2$, and the number of stages $M = 100$.  We select the best-fitting model using MDD and BIC. The results, summarized in Table~\ref{Tab:Iden}, show the good ability of both MDD and BIC to select the correct error term distribution, as well as the causal and noncausal orders. However, when such combinations of $P$, $M$, and $\lambda$ are considered, BIC outperforms MDD in detecting the correct model specification, especially when the error term follows a Skewed-$t$ distribution. In particular, under the Cauchy and Student-$t$ assumptions, the results show that the second most frequently identified model is the one with the correct error term distribution, but with overidentification of the causal order. In contrast, when Skewed-$t$ DGP is considered, the second most identified model not only shows a higher percentage than in the previous two cases, but is also associated with the wrong error term distribution (Student-$t$), despite correctly identifying the causal and noncausal orders. This suggests that due to the larger number of parameters associated with the skewness of the error term, more stages may be required to more closely track the evolution of the system. However, increasing $M$ becomes computationally impractical due to the significant increase in the Monte Carlo simulation time.

Table \ref{tab:est} displays the performance of the SMC estimates when polynomial orders and the error distribution are assumed to be known. The table displays the true parameter, the Monte Carlo average of the parameter estimates and their variance, the average of the standard errors obtained in each replication, the average bias (BIAS):
\begin{equation*}
   BIAS_{j}^{\mathcal{D}} = \dfrac{1}{B} \sum_{b=1}^B  \left(\widehat{\bTheta}^{\mathcal{D}}_{j,b}-\bTheta^{\mathcal{D}}_{j,0}\right),
\end{equation*}
and the average root mean squared error (RMSE):
\begin{equation*}
   RMSE_{j}^{\mathcal{D}} = \sqrt{\dfrac{1}{B}\sum_{b=1}^B \left(\widehat{\bTheta}^{\mathcal{D}}_{j,b}-\bTheta^{\mathcal{D}}_{j,0}\right)^2}.
\end{equation*}
$\widehat{\bTheta}^{\mathcal{D}}_{j,b}$ represents the estimated value, in simulation $b$ of the $j$-th element of the vector $\bTheta$, as defined in Section \ref{subsec:priors}, with dimension $N_\mathcal{D}$ changing according to the distribution of the error term, $\mathcal{D}$. Finally, $ \bTheta^{\mathcal{D}_i}_{j,0} $ denotes the true population value of the corresponding element $j$. The results show that when the error term follows a Cauchy distribution, the SMC algorithm performs well, with estimates of the parameters $\bPsi$, $\bPhi$, and $\Sigma $ being very close to the true values, as indicated by the small bias and the RMSE values. The variance of the estimates is also relatively small, suggesting a stable estimation between replications. 
For the Student-$t$ distribution, the parameter estimates show slightly larger deviations from the true values compared to the Cauchy case. This is reflected in the higher bias and the RMSE, but the algorithm still exhibits a good performance. Finally, for the Skewed-$t$ distribution, the SMC algorithm shows a slight negative bias for the scale and skewness parameters.

\

\begin{table}[H]
\centering
\caption{{The table displays the performance of the SMC algorithm in identifying the correct VMAR($1,1$) model under three multivariate error distributions: Cauchy, Student-$t$, and Skewed-$t$. The results are reported for two model selection criteria: MMD and BIC. Each row shows the percentage of times the algorithm correctly identified a particular model across different error distributions and polynomial orders. The table reports with (-) the model never selected, in the gray area the percentage of the most selected model, and in bold the second most selected model.}}
\resizebox{0.92\textwidth}{!}{%
\begin{tabular}{cccccccc}
\multicolumn{8}{c}{\textbf{Cauchy}}															\\	\midrule	
\multicolumn{8}{c}{MDD}															\\	\midrule	
	&	VMAR(1,0)	&	VMAR(0,1)	&	VMAR(2,0)	&	VMAR(1,1)	&	VMAR(0,2)	&	VMAR(2,1)	&	VMAR(1,2)	\\	\cmidrule(lr){2-8}	
Cauchy	&	0.5	&	-	&	-	&	\cellcolor{gray!20}\textbf{\textit{91.5}}		&	-	&	\textbf{5.0}	&	2.5	\\	
Student-$t$	&	-	&	-	&	-	&	0.5	&	-	&	-&	-	\\	
Skewed-$t$	&	-	&	-	&	-	&	-	&	-	&	-	&	-	\\	\midrule	
\multicolumn{8}{c}{BIC}															\\	\midrule	
Cauchy	&	-	&	-	&	-	&	\cellcolor{gray!20}\textbf{\textit{83.0}}	&	-	&	\textbf{10.0}	&	7.0	\\		
Student-$t$	&-		&	-	&	-	&		-&	-	&	-	&	-	\\		
Skewed-$t$	&	-	&	-	&	-	&	-	&	-	&	-	&	-	\\
\multicolumn{8}{c}{\textbf{Student-$\boldsymbol{t}$}}															\\	\midrule	
\multicolumn{8}{c}{MDD}															\\	\midrule	
Cauchy	&	-	&	-	&	-	&	1.5	&	-	&	-	&	0.5	\\		
Student-$t$	&	-	&	-	&	-	&	\cellcolor{gray!20}\textbf{\textit{85.5}}	&	-	&	\textbf{8.5}	&	3.5	\\	
Skewed-$t$	&	-	&	-	&	-	&	0.5	&	-	&	-	&	-	\\	\midrule	
\multicolumn{8}{c}{BIC}															\\	\midrule	
Cauchy	&	-	&	-	&	-	&	-	&	-	&	-	&	-	\\		
Student-$t$	&	-	&	-	&	1.5	&	\cellcolor{gray!20}\textbf{\textit{88.5}}	&	-	&	\textbf{6.5}	&	3.5	\\	
Skewed-$t$	&	-	&	-	&	-	&	-	&	-	&	-	&	-	\\		

\multicolumn{8}{c}{\textbf{Skewed-$t$}}															\\	\midrule	
\multicolumn{8}{c}{MDD}															\\	\midrule	
Cauchy	&	-	&	-	&	-	&	1.0	&	-	&	-	&	0.5	\\		
Student-$t$	&	-	&	-	&	-	&	\textbf{21.5}	&	-	&	0.5	&	-	\\	
Skewed-$t$	&	-	&	-	&	0.5	& 	\cellcolor{gray!20}\textbf{\textit{62.0}}	&	-	&	7.0	&	7.0	\\	\midrule	
\multicolumn{8}{c}{BIC}															\\	\midrule	
Cauchy	&	-	&	-	&	-	&	-	&	-	&	-	&	-	\\		
Student-$t$	&	-	&	-	&	-	&	0.5	&	-	&	-	&	-	\\		
Skewed-$t$	&	-	&	-	&	0.5	&	\cellcolor{gray!20}\textbf{\textit{94.0}}	&	-	&	2.0	&	\textbf{3.0}	\\	\bottomrule	
\end{tabular}}
\label{Tab:Iden}
\end{table}

\begin{table}[H]
\centering
\caption{The table displays the performance of the SMC algorithm in estimating the VMAR($1,1$) model when $r$, $s$, and the error distribution are assumed to be known, across three multivariate error distributions: Cauchy, Student-$t$, and Skewed-$t$. The results report the true parameter $\left(\bTheta_0\right)$, the Monte Carlo average of the parameter estimates $\left( \overline{\widehat{\bTheta}}\right)$ and their variance $\left(\text{Var}(\widehat{\bTheta})\right)$, the average of the standard errors obtained in each replication $\left(\overline{STD}\right)$, 
the average bias (BIAS), and the average Root Mean Squared Error (RMSE). Finally, $\bPsi_{ij}$, $\bPhi_{ij}$, and $\Sigma_{ij}$ are the $(i,j)$ elements of the causal, noncausal, and scale matrices, respectively, while $\nu$ denotes the degrees of freedom.}
\resizebox{0.85\width}{0.8\height}{%
\begin{tabular}{ccccccc}
\multicolumn{7}{c}{\textbf{DGP: Cauchy}}													\\	\midrule
	&	$\bTheta_0$	&	$\overline{\hat{\bTheta}}$	&	Var($\hat{\bTheta}$)	&	$\overline{STD}$	&	BIAS	&	RMSE	\\	\midrule
$\bPsi_{11}$	&	0.8000	&	0.8004	&	0.0001	&	0.0067	&	0.0004	&	0.0077	\\	
$\bPsi_{12}$	&	0.1000	&	0.1011	&	0.0002	&	0.0111	&	0.0011	&	0.0123	\\	
$\bPsi_{21}$	&	-0.2000	&	-0.2004	&	0.0001	&	0.0072	&	-0.0004	&	0.0077	\\	
$\bPsi_{22}$	&	0.3000	&	0.3002	&	0.0001	&	0.0102	&	0.0002	&	0.0094	\\	
$\bPhi_{11}$	&	0.6000	&	0.6008	&	0.0001	&	0.0073	&	0.0008	&	0.0085	\\	
$\bPhi_{12}$	&	-0.4000	&	-0.3992	&	0.0002	&	0.0114	&	0.0008	&	0.0130	\\	
$\bPhi_{21}$	&	-0.4000	&	-0.3994	&	0.0001	&	0.0058	&	0.0006	&	0.0074	\\	
$\Phi_{22}$	&	0.1000	&	0.1008	&	0.0002	&	0.0105	&	0.0008	&	0.0136	\\	
$\Sigma_{11}$	&	2.0000	&	2.1981	&	0.2357	&	0.5266	&	0.1981	&	0.5232	\\	
$\Sigma_{12}$	&	0.5000	&	0.4999	&	0.0676	&	0.2760	&	-0.0001	&	0.2593	\\	
$\Sigma_{22}$	&	2.0000	&	1.9488	&	0.3300	&	0.4662	&	-0.0512	&	0.5753	\\	\midrule
\multicolumn{7}{c}{\textbf{DGP: Student-$\boldsymbol{t}$}}													\\	\midrule
$\bPsi_{11}$	&	0.8000	&	0.7932	&	0.0024	&	0.0485	&	-0.0068	&	0.0496	\\	
$\bPsi_{12}$	&	0.1000	&	0.0941	&	0.0047	&	0.0721	&	-0.0059	&	0.0686	\\	
$\bPsi_{21}$	&	-0.2000	&	-0.1987	&	0.0026	&	0.0506	&	0.0013	&	0.0511	\\	
$\bPsi_{22}$	&	0.3000	&	0.2869	&	0.0053	&	0.0703	&	-0.0131	&	0.0741	\\	
$\bPhi_{11}$	&	0.6000	&	0.5866	&	0.0024	&	0.0514	&	-0.0134	&	0.0511	\\	
$\bPhi_{12}$	&	-0.4000	&	-0.3831	&	0.0046	&	0.0715	&	0.0169	&	0.0694	\\	
$\bPhi_{21}$	&	-0.4000	&	-0.3962	&	0.0012	&	0.0371	&	0.0038	&	0.0351	\\	
$\Phi_{22}$	&	0.1000	&	0.1058	&	0.0042	&	0.0647	&	0.0058	&	0.0649	\\	
$\Sigma_{11}$	&	2.0000	&	2.1616	&	0.1818	&	0.4351	&	0.1616	&	0.4550	\\	
$\Sigma_{12}$	&	0.5000	&	0.5153	&	0.0570	&	0.2670	&	0.0153	&	0.2387	\\	
$\Sigma_{22}$	&	2.0000	&	1.8847	&	0.1395	&	0.3683	&	-0.1153	&	0.3900	\\	
$\nu$	&	3.0000	&	3.2807	&	0.6337	&	0.7767	&	0.2807	&	0.8422	\\	\midrule
\multicolumn{7}{c}{\textbf{DGP: Skewed-$\boldsymbol{t}$}}													\\	\midrule
$\bPsi_{11}$	&	0.8000	&	0.7822	&	0.0021	&	0.0410	&	-0.0178	&	0.0488	\\	
$\bPsi_{12}$	&	0.1000	&	0.0945	&	0.0027	&	0.0576	&	-0.0055	&	0.0519	\\	
$\bPsi_{21}$	&	-0.2000	&	-0.1967	&	0.0025	&	0.0413	&	0.0033	&	0.0499	\\	
$\bPsi_{22}$	&	0.3000	&	0.2868	&	0.0069	&	0.0581	&	-0.0132	&	0.0841	\\	
$\bPhi_{11}$	&	0.6000	&	0.6213	&	0.0013	&	0.0367	&	0.0213	&	0.0418	\\	
$\bPhi_{12}$	&	-0.4000	&	-0.4090	&	0.0043	&	0.0604	&	-0.0090	&	0.0658	\\	
$\bPhi_{21}$	&	-0.4000	&	-0.3948	&	0.0017	&	0.0321	&	0.0052	&	0.0412	\\	
$\Phi_{22}$	&	0.1000	&	0.1053	&	0.0050	&	0.0608	&	0.0052	&	0.0710	\\	
$\Sigma_{11}$	&	2.0000	&	1.9373	&	0.2074	&	0.4551	&	-0.0627	&	0.4586	\\	
$\Sigma_{12}$	&	0.5000	&	0.2956	&	0.0534	&	0.2692	&	-0.2044	&	0.3080	\\	
$\Sigma_{22}$	&	2.0000	&	1.8499	&	0.1608	&	0.4365	&	-0.1501	&	0.4273	\\	
$\balpha_1$	&	2.0000	&	1.8936	&	0.2349	&	0.0117	&	-0.1064	&	0.4950	\\	
$\balpha_2$	&	2.0000	&	1.9559	&	0.2931	&	0.0143	&	-0.0441	&	0.4950	\\	
$\nu$	&	3.0000	&	3.3270	&	0.5363	&	0.7610	&	0.3270	&	0.8003	\\	
\bottomrule
\end{tabular}}
\label{tab:est}
\end{table}

\section{Empirical investigation}\label{sec:Empirical}
We illustrate the use of SMC in mixed models by considering a bivariate process involving the S$\&$P Europe 350 ESG Index and Brent crude oil prices. The S$\&$P Europe 350 ESG Index is a stock market index designed to measure the performance of companies in Europe that meet certain environmental, social, and governance (ESG) criteria. It is a variant of the broader S$\&$P Europe 350 Index, which tracks the largest and most liquid stocks in the European market. The goal here is to investigate the dynamic interactions between oil prices and the European ESG sector. The data, with monthly frequency, span from July 2017 to February 2024, comprising $T=116$ observations, and available at \url{https://www.spglobal.com/spdji/en/indices/sustainability/sp-europe-350-esg-index/#overview} and \url{https://fred.stlouisfed.org/series/DCOILBRENTEU}. 

\begin{figure}[!h]
     \centering
     \caption{\textbf{S$\boldsymbol{\&}$P Europe 350 Index and Brent prices}}
     \begin{subfigure}{0.45\textwidth}
         \includegraphics[width=7.5cm, height=4.5cm]{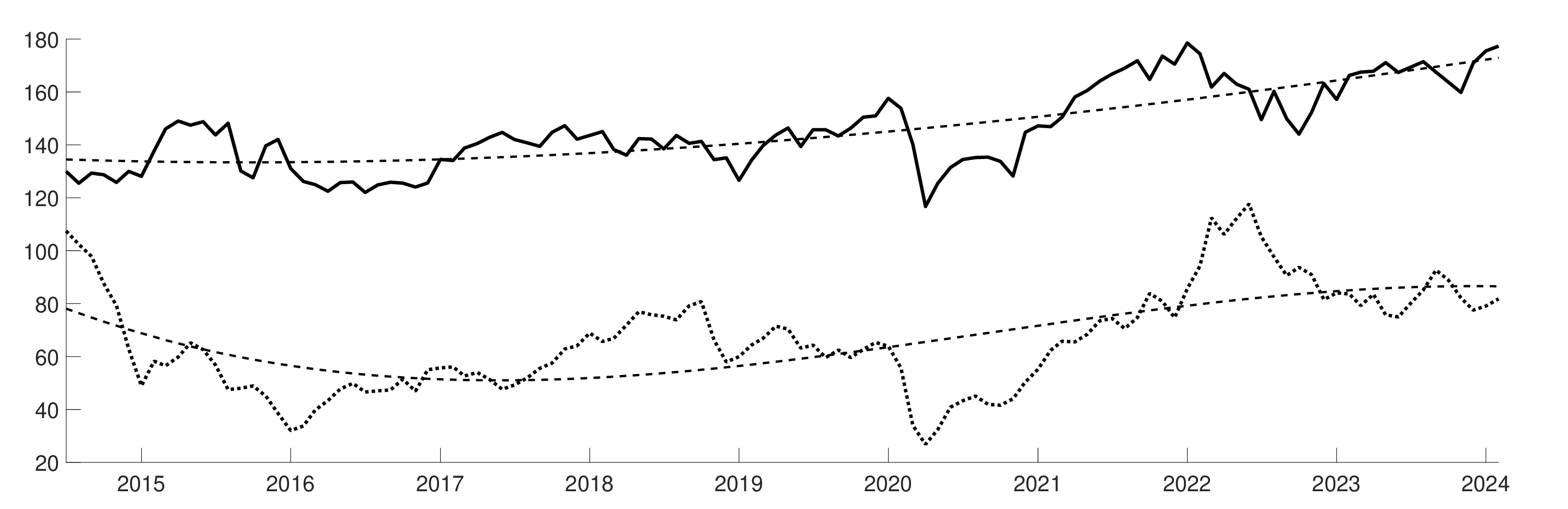}
    \caption{The graph displays the S\&P Europe 350 Index \\(solid black line) and Brent price (dashed line). The \\ trend component is obtained using a third-order \\ polynomial.}

     \end{subfigure}
     %\hfill
       \centering
     %\hspace{1.0cm}
     \begin{subfigure}{0.45\textwidth}
      \includegraphics[width=7.5cm,height=4.6cm]{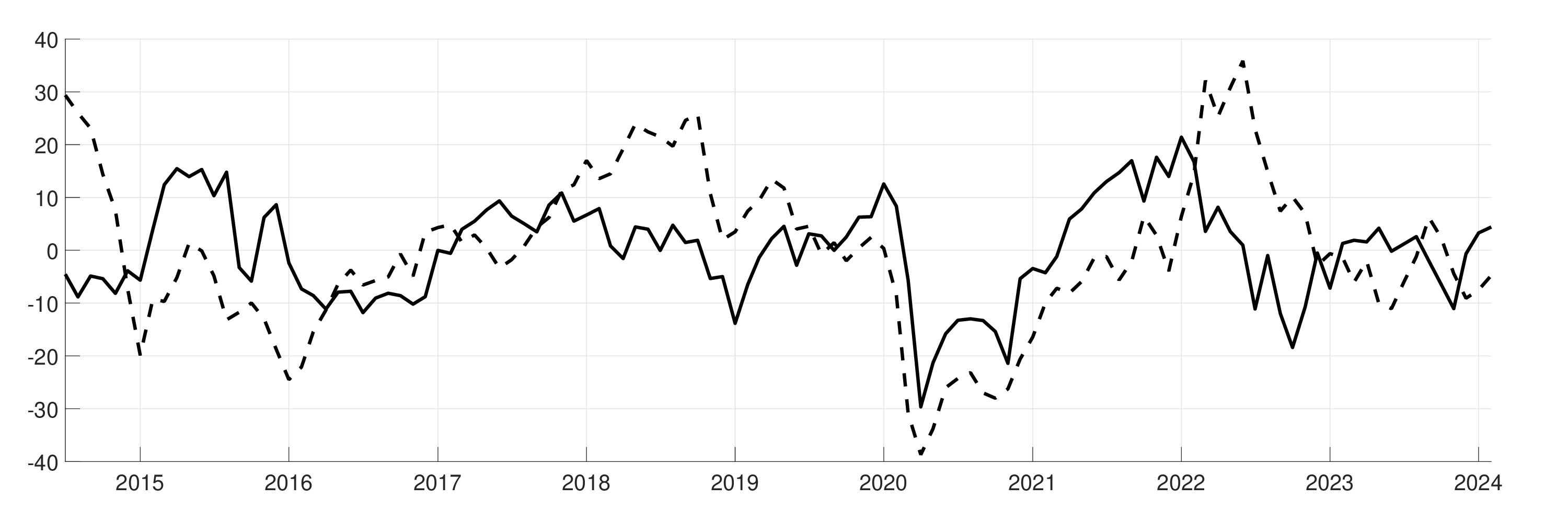}
        \caption{The graph displays the detrended series of\\ the S\&P Europe 350 Index (solid black line)\\ and Brent price (dashed line) using a third-order \\
        polynomial.}

     \end{subfigure}
     \label{fig:Emp}
\end{figure}

The time series are shown in Figure \ref{fig:Emp}-(a). As both series exhibit stochastic trends, we detrend them using a third-order polynomial. This choice strikes a balance between flexibility and parsimony, avoiding the risk of overfitting while capturing the underlying trend. Additionally, the third-order polynomial preserves key features in the data, such as the bubble patterns. Figure \ref{fig:Emp}-(b) displays the detrended series.

Next, using $P = 50000$, $M = 150$, and $\lambda = 2$, we estimate the series using the SMC algorithm, considering the following models: VMAR($1,0$), VMAR($0,1$), VMAR($2,0$), VMAR($1,1$), VMAR($0,2$), VMAR($2,1$), and VMAR($1,2$) under the Cauchy, Student-$t$, and Skewed-$t$ distributions. We select the error distribution and polynomial orders $r$ and $s$ that maximize MDD and minimize BIC (Table \ref{tab:Empiric}). Both methods suggest a VMAR$(1,1)$, with the Student-$t$ distribution providing the best fit for the error term. The estimated coefficients are shown in Table \ref{tab:Empiric2}.

\begin{table}[H]
\centering
\caption{{The table displays the performance of the SMC algorithm in identifying the bivariate process including the detrended S$\boldsymbol{\&}$P Europe 350 ESG Index and Brent Oil prices. The results are reported for two model selection criteria: Marginal Data Density (MDD) and Bayesian Information Criterion (BIC). In the grey area, the MDD and BIC better fit the data.}}
\resizebox{\textwidth}{!}{
\begin{tabular}{cccccccc}
\multicolumn{8}{c}{\textbf{Identification for the detrended series: (S$\boldsymbol{\&}$P,Brent)$^{\prime}$}}\\	\midrule
\multicolumn{8}{c}{MDD}															\\	\midrule
	&	VMAR(1,0)	&	VMAR(0,1)	&	VMAR(2,0)	&	VMAR(1,1)	&	VMAR(0,2)	&	VMAR(2,1)	&	VMAR(1,2)	\\	\cmidrule(lr){2-8}	
Cauchy	& -770.5 & -771.2 & -761.9 & -748.3 & -759.7 & -742.6 & -749.3 \\
Student-$t$	& -746.5 & -749.2 & -740.0 & \cellcolor{gray!20}\textbf{\textit{-728.9}} & -742.9 & -731.9 & -730.7 \\
Skewed-$t$	& -752.3 & -757.6 & -746.5 & -744.0 & -754.2 & -742.6 & -742.0 \\\midrule
\multicolumn{8}{c}{BIC}															\\	\midrule
Cauchy	& 1524.5 & 1521.6 & 1512.3 & 1492.9 & 1506.8 & 1468.5 & 1494.9 \\	
Student-$t$	& 1476.4 & 1521.6 & 1466.8 & \cellcolor{gray!20}\textbf{\textit{1457.9}} & 1470.5 & 1481.4 & 1458.2 \\
Skewed-$t$	& 1485.3 & 1500.5 & 1475.7 & 1481.3 & 1489.4 & 1478.2 & 1478.5 \\	
\bottomrule
\end{tabular}}
\label{tab:Empiric}
\end{table}												

\begin{table}[H]
\
\centering
\caption{The table displays the estimation of the bivariate VMAR($1,1$) process, involving the detrended S$\boldsymbol{\&}$P Europe 350 ESG Index and Brent Oil prices. The error term is assumed to follow a Student-$t$ distribution.\\}
\begin{tabular}{cccccc}
\multicolumn{6}{c}{\textbf{(S}{$\boldsymbol{\&}$\textbf{P,Brent)}$^{\prime}$}}\\	\midrule

& \multicolumn{2}{c}{r=1} && \multicolumn{2}{c}{s=1}\\  
\cmidrule(lr){2-3} \cmidrule(lr){5-6}          
$\bPsi_r$	&	0.32    &    -0.23  && -	& -	  \\
 %           &(0.18)	    &	(0.10)	&& 	   &      \\	
	   &	-0.17	&	0.70	 && -	& -    \\	
%	    &	(0.11)	&	(0.08)	&& 	   & 	  \\	
        &&&&&\\
$\bPhi_s$	&	-	    &	-	&& 0.50	& 0.21	\\	
%	&		    &		    && (0.20)	& (0.13)	 \\	
	&	-	    &	-	    && 0.43	& 0.51		\\	
%	&		    &		    && (0.14)	& (0.10) \\	
\midrule
$\Sigma$&	18.42	&	1.25    &&	-	& - \\
%	&	(3.50)	&	(2.12)	&&		&		 \\	
	&	1.25	&	14.93	&&	-	& - \\	
%	&	(2.12)	&	(3.55)	&&		&		 \\
\midrule
$\nu$	&	3.99	&	-	&&	-	&	- \\	
%	&	(1.06)	&	-	&&		&	 \\	
\bottomrule																	
\end{tabular}
\label{tab:Empiric2}
\end{table}

The 0.43 coefficient in matrix $\bPhi_1$ shows that an expected future increase of S$\boldsymbol{\&}$P Europe 350 ESG Index tends to increase the Brent price. A possible interpretation of the results may lie in the fact that as companies improve their ESG profiles, they may invest in greener technologies or renewable energy sources, which can lead to an increase in the short-term demand for oil. In other words, companies might need oil during the transition phase toward sustainability, where they are still reliant on fossil fuels for energy. Thus, a positive shift in ESG perceptions can create a paradoxical increase in demand for Brent, as firms balance immediate operational needs with long-term sustainability goals. Furthermore, Table \ref{tab:Empiric2} also suggests that an expected increase in Brent prices slightly increases the S$\&$P Europe 350 ESG Index, implying that higher oil prices may positively impact the stock performance of ESG-compliant companies in Europe.

\section{Conclusions}\label{sec:Concl}
The paper extends and adapts the Bayesian SMC algorithm to handle MAR processes. Key advantages of the SMC method include faster initialization, parallelized computations, mitigation of issues related to local minima, and automatic generation of Marginal Data Density (MDD) as a byproduct. Consequently, SMC can be easily applied to a wide range of error term distributions, provided that the MDD can be estimated point-wise. Moreover, since the distribution of the error term is typically unknown to researchers, we exploit the adaptability of SMC to propose a novel identification procedure. This procedure, based on MDD and the BIC, differs from previous studies in that it not only selects the causal and noncausal polynomial orders but also identifies the error term distribution that best fits the data. As a result, we explore the Bayesian estimation of such models for the first time, not only in the Student-$t$ framework. Monte Carlo experiments demonstrate that SMC performs well in terms of bias and RMSE, successfully identifying the causal and noncausal polynomial orders as well as the error term distribution.

In the empirical application, we applied the SMC algorithm to a 2-dimensional VMAR($r,s$) model, including the S$\&$P Europe 350 ESG index and Brent crude oil prices. The causal and noncausal polynomial orders of the bivariate process are detected as $r=s=1$. Furthermore, we find that the assumption of multivariate Student-$t$ distribution is the one that better fits the data compared to the Cauchy and Skewed-$t$ distributions. Finally, our results indicate that an expected increase in the S$\&$P Europe 350 ESG Index can lead to a higher demand for Brent crude oil, while an expected rise in Brent prices positively influences the S$\&$P Europe 350 ESG Index.

\newpage
\appendix
\section{Appendix}
\label{appendix:univariate}
When considering the MAR($r,s$) process, we define 
$\psi=\left(\psi_1, \dots, \psi_r\right)^{\prime}$ 
and $\phi=\left(\phi_1, \dots, \phi_r\right)^{\prime}$.
Furthermore, $\sigma$, $\nu$, and $\alpha$ represent the scale parameter, the degrees of freedom, and the skewness parameter, respectively.

\noindent\textbf{Priors}\\
As discussed in Section \ref{subsec:BayesianVMAR}, $p(\theta)$ represents the prior probability of the MAR parameters, and $\theta=\left[\theta_1, \theta_2 \right]$. Specifically, $\theta_1 = \left[\psi^{\prime},\phi^{\prime} \right]^{\prime}$ remains unchanged regardless of the error term distribution. In contrast, $\theta_2$ varies with the density distribution of $u_t$. This vector includes the scale parameter $\sigma$, which is shared by the Student-$t$, Cauchy, and Skewed-$t$ distributions; the degrees of freedom ($\nu$) for the Student-$t$ and Skewed-$t$ distributions; and the vector containing the skewness parameters ($\alpha$) for the Skewed-$t$ distribution.

For $\psi$ and $\phi$ we use Multivariate Normal priors:
\begin{equation*}
    \psi \sim \mathcal{MN}\left(0, \gamma \text{I}_{r}\right)\calI(\psi),
        \hspace{1.5cm}  \phi \sim \mathcal{MN}\left(0, \delta \text{I}_{s}\right)\calI(\phi),
\end{equation*}
where: I$_{r}$ and I$_{s}$ are identity matrices of dimension $r \times r$ and $s\times s$, respectively; and $\calI(\psi)$ and $\calI(\phi)$ are indicator functions equal to 1 in the stationary regions. Furthermore, we set $\gamma=2/i$, for $i=1, \dots , r$ in the causal component and $\delta=2/q$, for $q=1, \dots , s$ in the noncausal one. The prior distribution for the scale matrix, which is common to all considered distributions, is the following non-informative standard prior distribution:
\begin{equation*}
    p(\omega) \propto 1/\sigma,
\end{equation*}
where $\omega=\sigma^{-2}$. Finally, we use Exponential prior ($\mathcal{E}(\cdot)$) for the degrees of freedom ($\nu$) and $\mathcal{N}(\cdot, \cdot)$ for the skewness parameter $\alpha$:
\begin{equation*}
    \nu \sim \mathcal{E}(\nu_0),  \hspace{2cm}  \alpha \sim \mathcal{N}(0,\kappa),
    \label{eq:df}
\end{equation*}
where, we set $\nu_0=5$ and $\kappa=3$.

\noindent\textbf{Likelihood functions}\\
The second component of the Bayesian analysis described in Section \ref{subsec:BayesianVMAR} is the likelihood function $p(y \mid \theta)$. We use the same (approximate) likelihood function used in \cite{lanne2011noncausal}, given by:
\begin{equation*}
    p(y \mid \theta)=\sum_{t=r+1}^{T-s}p(u_t \mid \theta),
\end{equation*}
where $T$ is the sample size, and $p(u_t \mid \theta)$ varies according to the error term distribution considered. Specifically, for the Student-$t$ we have:
\begin{equation}
    p\left(u^{Student-t}_t \mid \theta \right) = \frac{\Gamma\left(\frac{\nu + 1}{2}\right)}{\Gamma\left(\frac{\nu}{2}\right) \sqrt{\nu \pi \sigma^2}} \left(1 + \frac{\left[\psi(L)\phi(L^{-1})y_t\right]^2}{\nu \sigma^2}\right)^{-\frac{\nu + 1}{2}}.
    \label{pr:student}
\end{equation}
For the Skewed-$t$, we have:
\begin{equation}
	p\left(u^{Skewed-t}_t \mid \theta \right)= \frac{2}{\alpha+ \frac{1}{\alpha}}\left[ t\left(\frac{u_t}{\alpha} \right)
	\mathcal{I}(u_t) + t(\alpha u_t) \mathcal{I}(- u_t) \right],
    \label{pr:skewed}
\end{equation} 
where $\mathcal{I} (u_t)$ and $\mathcal{I} (-u_t)$ stand for the indicator function:
\begin{center}
$\mathcal{I} (u_t)=
\begin{cases}
1, \ \ \ u_t \geq 0 \\
0, \ \ \ u_t < 0\\
\end{cases},$
\end{center}
$t(u_t)$ stands for the density function of a symmetric Student-$t$, and $\alpha \in \mathbb{R}^{+}$. In case $\alpha = 1$, then \eqref{pr:student} and \eqref{pr:skewed} coincide (see \citealp{giancaterini2022climate}). Finally, when the Cauchy distribution is considered, then $p^{Cauchy}(u_t \mid \theta)$ is as in \eqref{pr:student}, with $\nu=1$; see \cite{gourieroux2017local}.

\section{Appendix}
\label{appenidix_B}
As seen in \cite{bognanni2018sequential}, the SMC algorithm consists of three steps, which will be presented here. The algorithm starts by drawing particles from $p(\bTheta)$ and giving them equal weights. Then, the recursive process begins. At each stage $m$, we have an approximation of the particles $\{\bTheta_{m-1}, W^i_{m-1}\}_{i=1}^{P}$ to $\pi_{m-1}$. The first step of stage $m$, the correction step, involves reweighting the particles according to an importance sampling of $\pi_m$ using $\pi_{m-1}$ as the source distribution. The purpose of this correction is to adjust the particle weights so that they better represent the target distribution $\pi_m$. In the second step, known as the selection step, the particles are rejuvenated by using multinomial resampling when the sample exhibits an imbalance, which means that only a few particles have meaningful weights. However, resampling introduces noise into the simulation, so it is only performed when necessary. The last stage, known as the mutation step, involves applying $S$ iterations of an MCMC algorithm (with invariant distribution $\pi_m$) to move particles in the parameter space. This step is essential to shift the particles to areas with a higher density of $\pi_m$ and to ensure diversity after the selection step. Without mutation, repeated resampling would leave only a few unique values, resulting in a poor posterior approximation. However, the correction step ensures that the particle distribution is approximately $\pi_m$ before mutation, allowing the procedure to be valid even for short chains such as $M = 1$.

\begin{algorithm}[H]
\caption{Sequential Monte Carlo (SMC)}\label{alg:your_algorithm}
\begin{algorithmic}
\small % Reduce font size
\STATE \textbf{Initialization} ($\rho_1=0$): Draw the initial particles from the prior distribution $\bTheta_{i}^{i.i.d.}\sim p(\bTheta)$ and assign them uniform weights equal to 1.
\begin{equation*}
    \bTheta_1^{i^{\text{ i.i.d.}}} \sim p(\bTheta), \ \ W_1^i = 1, \ \ i=1, \dots , P,
\end{equation*}
\FOR{$m=2$ \TO $M$}
\STATE \textbf{1. Correction Step}: Adjust the weights of the particles from stage $(j-1)$ by establishing the incremental and normalized weights:
  \begin{equation*}
      \tilde{w}_m^i = [ p(\by|\bTheta_{m-1}^i)]^{\rho_m-\rho_{m-1}}, \ \ \ \tilde{W}_m^i = \frac{\tilde{w}_m^i \tilde{W}_{m-1}^{i}}{\frac{1}{P}\sum_{i=1}^P \tilde{w}_m^i \tilde{W}_{m-1}^{i}}, \ \ \ i=1,\dots,P,
  \end{equation*}
\STATE \textbf{2. Selection}: Calculate the effective sample size:
  \begin{equation*}
      ESS_m = P/ \left( \frac{1}{P} \sum_{i=1}^{P} \left(\tilde{W}_m^i\right)^2\right),
  \end{equation*}
\IF{$ESS_m < \frac{P}{2}$}
    \STATE Resample the particles via multinomial resampling and reinitialize the weights to uniform:
    \begin{equation*}
        W_m^i = 1, \ \ \ \widehat{\bTheta}_m^i \sim \{\bTheta_{m-1}^{p}, \tilde{W}_m^p \}_{p=1, \dots, P}, \ \ i = 1, \dots, P,
    \end{equation*}
\ELSE
    \STATE 
    \begin{equation*}
            W_m^i = \tilde{W}_m^i, \ \ \ \widehat{\bTheta}_m^i = \bTheta_{m-1}^{i}, \ \ \ i = 1, \dots ,P,
    \end{equation*}
\ENDIF \\
\STATE \textbf{3. Mutation}: Propagate each particle $\{\widehat{\bTheta}_m^i, W_m^i \}$ via $S$ steps of an MCMC algorithm with transition density $\bTheta_m^i \sim K_m(\bTheta_m|\widehat{\bTheta}_m^i; \zeta_m)$ and stationary distribution $\pi_m(\bTheta)$. See \cite{bognanni2018sequential} for details on the transition density.
\ENDFOR \\

\

\STATE \textbf{Compute posterior moments}: An approximation of $E_{\pi_m} [ h(\bTheta)]$ is given by
\begin{equation*}
    \bar{h}_{m,P} = \frac{1}{P}\sum_{i=1}^{P} h(\bTheta_m^i) W_m^i,
\end{equation*}
\STATE This approximation is valid using the particle approximations $\{\bTheta_{m-1}^i, \tilde{W}_m^i \}_{i=1}^P$, $\{\widehat{\bTheta}_{m}^i, \tilde{W}_m^i \}_{i=1}^P$, and $\{\bTheta_m^i, W_m^i \}_{i=1}^{P}$ after the correction, selection, and mutation steps, respectively. \\

\

\STATE \textbf{Estimate of marginal data density}: An estimate of the marginal data density is given by
\begin{equation*}
    \widehat{p(Y)} = \prod_{m=1}^{M} \Biggl( \sum_{i=1}^{P} \tilde{w}_m^i \Biggr). 
\end{equation*}
\end{algorithmic}
\end{algorithm}

\newpage

\bibliographystyle{chicago}
\bibliography{references}
%\printbibliography 
\end{document}